\documentclass[12pt,a4paper,aps,preprint,superscriptaddress,nofootinbib]{revtex4-1}
\usepackage{graphicx}
\usepackage{amssymb}
\usepackage{slashed}
\usepackage{textcomp}
\usepackage{amsmath}
\usepackage{bm}
\usepackage{times}
\usepackage{epsfig}
\usepackage{color}
\usepackage{ulem}
\usepackage{subfigure}
\usepackage{ulem}

\newcommand{\be}{\begin{equation}}
\newcommand{\ee}{\end{equation}}
\newcommand{\ba}{\begin{array}}
\newcommand{\ea}{\end{array}}
\newcommand{\bea}{\begin{eqnarray}}
\newcommand{\eea}{\end{eqnarray}}
\newcommand{\nn}{\nonumber}

\def\lsim{\mathrel{\raise.3ex\hbox{$<$\kern-.75em\lower1ex\hbox{$\sim$}}}}
\def\gsim{\mathrel{\raise.3ex\hbox{$>$\kern-.75em\lower1ex\hbox{$\sim$}}}}
\usepackage{ulem,fancyvrb}
\usepackage{xcolor}

\makeatother

\begin{document}
	
\title{ Exploring dark matter-gauge boson effective interactions at the current and future colliders}

\author{Yu Zhang}
\affiliation{Institutes of Physical Science and Information Technology,
	Anhui University, Hefei 230601, China  }
\affiliation{School of Physics, Hefei University of Technology, Hefei 230601,China}
\author{Yi-Wei Huang}
\affiliation{Institutes of Physical Science and Information Technology,
	Anhui University, Hefei 230601, China  }
\author{Wei-Tao Zhang}
\affiliation{School of Physics and Optoelectronics Engineering, Anhui University, Hefei 230601,China}
\author{Mao Song}
\affiliation{School of Physics and Optoelectronics Engineering, Anhui University, Hefei 230601,China}
\author{Ran Ding}
\email{Corresponding author: dingran@mail.nankai.edu.cn}
\affiliation{School of Physics and Optoelectronics Engineering, Anhui University, Hefei 230601,China}


\date{\today}

\begin{abstract}
	In this work, we investigate the collider constraints on effective interactions between Dark Matter (DM) particles and electroweak gauge bosons in a systematic way. We consider the simplified models in which scalar or Dirac fermion DM candidates only couple to electroweak gauge bosons through high dimensional effective operators. Taking into account the induced DM-quarks and DM-gluons operators from the Renormalization Group Evolution (RGE) running effect, we present comprehensive constraints on effective energy scale $\Lambda$ and Wilson coefficients $C_B(\Lambda),\,C_W(\Lambda)$ from direct detection, indirect detection, and collider searches. In particular, we present the corresponding sensitivity from the Large Hadron Electron Collider (LHeC) and the Future Circular Collider in electron-proton mode (FCC-ep) for the first time, update the mono-$j$ and mono-$\gamma$ search limits at the Large Hadron Collider (LHC), and derive the new limits at the Circular Electron Positron Collider (CEPC).
\end{abstract}	
	
\maketitle

\section{Introduction}\label{sec:intro}

Observations from astrophysics and cosmology have provided overwhelming evidence for the existence of DM~\cite{Aghanim:2018eyx}. However, the microscopic properties of DM are still poorly known since all of such observations are based on its gravitational effects. The canonical DM candidate is the Weakly Interacting Massive Particles (WIMPs). Such a scenario is easy to fulfill relic abundance through thermal freeze-out mechanism (the ``WIMP miracle'')~\cite{Lee:1977ua}, and can be naturally embedded in many popular theoretical frameworks~\cite{Jungman:1995df,Bertone:2004pz}. Moreover, due to the property of weak interaction with the SM particles, it can be probed through three experimental prongs, i.e., collider, direct and indirect detections.

Typically, the interactions between DM and SM particles are model-dependent and it is a great challenge to explore the complete model landscape. Instead of exhausting all possible theoretical model parameter space, the Effective Field Theory (EFT) framework is extensively used, which allows us to capture some key features of the high-scale physics effects on the electroweak scale while significantly simplifying the analysis procedures. For EFT collider searches, various signal channels based on different SM portal effective operators have been extensively studied at the hadron and electron colliders~\cite{Goodman:2010ku,Fox:2011pm,Rajaraman:2011wf,Ding:2013nvx,Cao:2009uw,Nelson:2013pqa,Crivellin:2015wva,Bell:2012rg,Carpenter:2012rg,Alves:2015dya,Mao:2014rga,Lopez:2014qja,Carpenter:2013xra,Berlin:2014cfa,Bartels:2012ex,Chae:2012bq,Dreiner:2012xm,Yu:2013aca,Richard:2014vfa,Rossi-Torres:2015eua,Habermehl:2017dxh,Gao:2017tgx,Liu:2019ogn,Ghosh:2019rtj,Habermehl:2020njb,Neng:2014mga,Yu:2014ula,Liu:2017zdh,Kadota:2018lrt,dEnterria:2017dac}.

Among them, the parameter space of DM-quark interaction which fulfills the correct relic density has a tension with the exclusion limits from direct detection, indirect detection, and collider searches. In this work, we consider the dimension-6 (7) DM-diboson effective operators for scalar (Dirac fermion) DM and examine their sensitivity at current LHC, future electron and electron-proton colliders. The current limits of DM-diboson operators at the LHC are mainly constrained by mono-$j$ and mono-$\gamma$ signal channels, and mono-$\gamma$ signature is a sensitive channel at the CEPC. For electron-proton collider,  LHeC and FCC-ep are the two most promising proposals at current. LHeC is based on an economic LHC upgrade, plan to add a 60-140 GeV electron beam to collide with a 7 TeV proton beam in the LHC ring during the period of High Luminosity LHC (HL-LHC) run. FCC-ep is designed to collide a 60 GeV electron with the 50 TeV proton beam. DM pairs can be produced through the VBF process with a cleaner background, which corresponds to ${\slashed E}_T+e^-j$ signal.

The rest paper is organized as follows. In Sec.\ref{sec:eft}, we firstly give a brief description of the DM-diboson effective operators. Then relevant collider phenomenology of ${\slashed E}_T+e^-j$ signal channel at the LHeC and FCC-ep is discussed in Sec.\ref{sec:lhec}, including event simulations, kinematical distributions of final states and cut selections related to signals and backgrounds. In Sec.\ref{sec:cepc}, we present the sensitivity at the CEPC with mono-$\gamma$ signature. The limits from the LHC with current mono-$j$ and mono-$\gamma$ searches are updated in Sec. \ref{sec:LHC}. The constraints from current direct and indirect DM search experiments are also discussed in Sec.\ref{sec:direct}. Then we show the results in Sec. \ref{sec:results}. Finally, we give a summary and conclusion in Sec.\ref{sec:summary}.

\section{Framework of DM-gauge boson effective interactions}\label{sec:eft}

We consider the DM particle as a Dirac fermion ($\chi$) or scalar ($\phi$), which connects with SM sector through high-dimensional DM-gauge boson effective operators. At the lowest order, such operators allow DM couple to single gauge boson, which induce electromagnetic dipole and anapole interactions~\cite{Cotta:2012nj, Rajaraman:2012fu}. However, these interactions lead to monochromatic gamma-rays due to the unsuppressed cross-section of DM annihilating into the $\gamma\gamma$ and $\gamma Z$ modes, which are severely constrained by current indirect detection experiments ~\cite{Chen:2013gya}. We thus start from DM-diboson operators
\bea
{\cal O}_B^\phi &=& \phi^*\phi B^{\mu\nu}B_{\mu\nu}, \qquad {\cal O}_W^\phi = \phi^*\phi W^{a,\mu\nu}W^a_{\mu\nu}, \label{eq:ppvv}\\
{\cal O}_B^\chi &=& \overline{\chi}\chi B^{\mu\nu}B_{\mu\nu}, \qquad
{\cal O}_W^\chi = \overline{\chi}\chi W^{a,\mu\nu}W^a_{\mu\nu}. \label{eq:ccvv}
\eea
These operators are assumed to be generated at the energy scale $\Lambda$ with other high energy resonances much heavier than $\Lambda$ are all decoupled. Notice that in this scenario, the coupling between DM and the SM quarks can also be induced through the renormalization group evolution (RGE) from the $\Lambda$ scale down to the electroweak scale $\mu_{\rm EW}$. As a consequence, two extra dimension-6 (7) operators
\bea
{\cal O}_y^\phi &=& y_q\phi^*\phi \bar q H q, \qquad {\cal O}_H^\phi = \phi^*\phi(H^\dagger H)^2,  \\
{\cal O}_y^\chi &=& y_q\bar \chi \chi \bar q H q, \qquad {\cal O}_H^\phi = \bar \chi \chi (H^\dagger H)^2.
\eea
need to be taken into account. In above equation, $y_q=\sqrt{2}m_q/v$ are the Yukawa couplings of SM quarks, and $v$ is the vacuum expectation value (VEV) of the SM Higgs doublet $H$. Then the total effective Lagrangian at scale $\mu$ ($\mu_{\rm EW} < \mu < \Lambda$) is given by
\begin{align}
\mathcal{L}_{\rm eff}^{\phi,\chi} = \sum_{k=B,W,y,H} \frac{C^{\phi,\chi}_k(\mu)}{\Lambda^2} {\cal O}_k\,.
\label{eq:lag}
\end{align}
To the leading logarithmic (LL) order, the Wilson coefficients of the effective operators ${\cal O}_y$ and ${\cal O}_H$ at the scale $\mu_{\rm EW}$ are~\cite{Crivellin:2014gpa}
\bea
C_y(\mu_{\rm EW}) &\simeq& \frac{6Y_{qL}Y_{qR}\alpha_1}{\pi}\ln\left(\frac{\mu_{\rm EW}}{\Lambda}\right), \\
C_H(\mu_{\rm EW}) &\simeq& -9 \alpha^2_2\ln\left(\frac{\mu_{\rm EW}}{\Lambda}\right) C_W(\Lambda),
\eea
where $Y_{qL}\,(Y_{qR})$ denotes the hypercharges of the left-handed (right-handed) quarks
with $Y_{uL}=Y_{dL}=1/6, Y_{uR}=2/3$ and $Y_{dR}=1/3$, $\alpha_1$ and $\alpha_2$ are the
gauge coupling constants of the $U(1)_Y$ and $SU(2)_Y$ gauge group, respectively. In our calculation, we set
$\mu_{\rm EW}\equiv m_Z$ with $\alpha_1\simeq 1/98$ and $\alpha_2\simeq 1/29$. After Electroweak symmetry breaking (EWSB), the gauge eigenstate fields $B_\mu$ and $W_\mu^3$ mix into the physical massless photon field $A_\mu$ and massive gauge field $Z_\mu$. Then the effective DM-diboson and DM-quark operators are recast as
\begin{align}
{\cal O}_{FF}^\phi &= \phi^*\phi F^{\mu\nu}F_{\mu\nu}, \quad {\cal O}_{FF}^\chi = \overline{\chi}\chi F^{\mu\nu}F_{\mu\nu}, \nonumber\\
{\cal O}_{q}^\phi &= m_q \phi^*\phi\bar{q}q, \quad {\cal O}_{q}^\chi = m_q \overline{\chi}\chi\bar{q} q.
\end{align}
where $FF=AA,AZ,ZZ,WW$. The relevant matching conditions are
\begin{eqnarray}
{C_{AA}(\mu_{\rm EW})} &=&C_B(\mu_{\rm EW}) \cos ^{2} \theta_{W}+C_W(\mu_{\rm EW}) \sin ^{2} \theta_{W}, \nn\\
{C_{Z Z}(\mu_{\rm EW})} &=&C_W(\mu_{\rm EW}) \cos ^{2} \theta_{W}+C_B(\mu_{\rm EW}) \sin ^{2} \theta_{W},  \nn\\
{C_{A Z}(\mu_{\rm EW})} &=&\left(C_W(\mu_{\rm EW})-C_B(\mu_{\rm EW})\right) \sin 2 \theta_{W}, \nn\\
{C_{W W}(\mu_{\rm EW})}&=&C_W(\mu_{\rm EW}), \nn\\
{C_{q}(\mu_{\rm EW})}&=&C_y(\mu_{\rm EW})-\frac{v^2}{m_h^2}C_H(\mu_{\rm EW}),
\label{eq:vortex}
\end{eqnarray}
with $\theta_{W}$ is the Weinberg angle.

Integrating out the heavy top quarks, the effective DM-gluon operators
\begin{align}
{\cal O}_{q}^\phi = \alpha_s \phi^*\phi G^{a,\mu\nu}G_{\mu\nu}^a, \quad {\cal O}_{q}^\chi = \alpha_s \overline{\chi}\chi G^{a,\mu\nu}G_{\mu\nu}^a
\end{align}
can be generated, where $G_{\mu\nu}^a$ denotes the gluon field strength. The matching condition at leading order is given by \cite{Shifman:1978zn}
\begin{align}
C_{G}(m_t)=-\frac{1}{12\pi}C_t(m_t).
\label{eq:cg}
\end{align}

When RGEs are evolved from the EW scale down to the hadronic scale $\mu_{\rm had}\simeq 1$ GeV, ${\cal O}_{AA}$ will contribute to ${\cal O}_{q}$ through the exchanging of virtual photons.
For scale $\mu$ with $m_q<\mu<\mu_{\rm EW}$, one has~\cite{Frandsen:2012db}
\be
{C_{q}(\mu)}\simeq C_{q}(\mu_{\rm EW})+ \frac{6 Q_q^2\alpha_{\rm em}}{\pi} \ln\left(\frac{\mu}{\mu_{\rm EW}}\right) C_{AA}(\mu_{\rm EW})
\ee
at LL order, where $Q_q$ is the electric charge of quark and the electromagnetic coupling constant $\alpha_{\rm em} \simeq 1/137$.

At $m_b$ and $m_c$ threshold, one has to integrate out the corresponding heavy bottom and charm quarks
by taking into account Eq.~(\ref{eq:cg}) again. Now at scale $\mu_{\rm had}\simeq 1$ GeV, we get
\bea
C_{q}\left(\mu_{\rm had}\right) &\simeq& \left(\frac{6 Y_{q_{L}} Y_{q_{R}} \alpha_{1}}{\pi} C_{B}(\Lambda)+ 9 \alpha_2^2 \frac{v^{2}}{m_{h}^{2}} C_{W}(\Lambda)\right) \ln \left(\frac{\mu_{\rm EW}}{\Lambda}\right)\nonumber \\
&+&\frac{6 Q_{q}^{2} \alpha_{\rm em}}{\pi} C_{AA}(\mu_{\rm EW}) \ln \left(\frac{\mu_{\rm had}}{\mu_{\rm EW}}\right), \\
C_{G}\left(\mu_{\rm had}\right) &\simeq& -\frac{1}{12 \pi}\Bigg\{\left(\frac{\alpha_{1}}{\pi} C_{B}(\Lambda)+ 27 \alpha_2^2 \frac{v^{2}}{m_{h}^{2}} C_{W}(\Lambda)\right) \ln \left(\frac{\mu_{\rm EW}}{\Lambda}\right) \nonumber \\
&+&\frac{2 \alpha_{\rm em}}{3 \pi} C_{AA}(\mu_{\rm EW})\left[\ln \left(\frac{m_b}{\mu_{\rm EW}}\right)+4 \ln \left(\frac{m_c}{\mu_{\rm EW}}\right)\right]\Bigg\},
\eea
where the Wilson coefficient $C_{AA}(\mu_{\rm EW})$ is determined by the first matching condition in Eq.~(\ref{eq:vortex}).

\section{Sensitivities at the LHeC and FCC-ep }\label{sec:lhec}

In this section, we will investigate the sensitivity of the DM-gauge boson effective operators at future electron-proton colliders.

\subsection{Signal channels and SM backgrounds}

\begin{figure}[htbp]
	\vspace{0.2cm}
	\centering
	\includegraphics[scale=0.8]{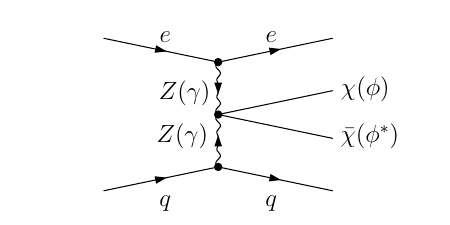}
	\caption{Feynman diagrams for the processes $e^- p \to e^-j \chi\bar\chi(\phi\phi^*)$}	
	\label{fig:process}
\end{figure}

From the effective interactions in Eqs.~(\ref{eq:ppvv}) and ~(\ref{eq:ccvv}), DM pairs can be produced at the electron-proton collider through both the Charged Current (CC) and the Neutral Current (NC), which respectively correspond to $W$ boson mode and $\gamma/Z$ mode of VBF processes. In CC production, DM can be generated through the process of $e^- p \to \nu_e j \chi\bar\chi(\phi\phi^*)$ via $WW$ fusion, resulting in a mono-jet plus missing energy signature. It is important to note that the mono-jet signature coincidentally aligns with the background of CC deeply inelastic scattering. Due to the absence of kinematic handles in the final state, distinguishing this signal channel from its primary background poses a significant challenge. Consequently, our study is primarily focused on NC production - specifically the process $e^- p \to e^-j \chi\bar\chi~(\phi\phi^*)$, as depicted in Fig.~\ref{fig:process}. The corresponding signature is thereby characterized as the ${\slashed E}_T+e^-j$ channel.

For SM background estimation, we consider both reducible background (RB) and irreducible background (IB). The IB mainly is due to the processes $e^-p\to e^-j\nu_\ell\bar\nu_\ell~(\ell=e,~\mu,~\tau)$. Among them, both $W/Z$-boson bremsstrahlung processes $e^-p\to e^-jZ/\nu_ejW^-~(Z\to\nu_e\bar\nu_e/W^-\to e^-\bar\nu_e)$ contribute to $\nu_e$ final state, while for $\nu_{\mu,\tau}$,
only bremsstrahlung process $e^-p\to e^-jZ~(Z\to\nu_{\mu,\tau}\bar\nu_{\mu,\tau})$ is relevant. For RB, a major contribution comes from $e^- p \to e^- j W^\pm$ with $W$'s leptonic decay $W^\pm\to\ell^\pm \nu_\ell$. There then two possibilities that need to be taken into account. One is that electron and muon final states run out of the detector acceptance and $\tau$ final state fakes a hard jet in the detector. Another possibility is that the products of hadronic $\tau$ decays are too soft to be tagged \cite{Tang:2015uha,Han:2018rkz}. As a consequence, The SM background can be summarized as follows:
\begin{itemize}
	\item  IB: $e^-p\to e^-j\nu_\ell\bar\nu_\ell$,
	\item  RB: $e^- p \to e^- j \ell^\pm \nu_\ell~(\bar\nu_\ell)$,
\end{itemize}
where $\ell=e,\mu,\tau$.

\subsection{Event simulation and kinematical cuts}

For the purpose of event simulation, we generate Universal FeynRules Output (UFO) \cite{Degrande:2011ua} model file by using {\sc FeynRules}~\cite{Alloul:2013bka}.  Which is fed into {\sc MadGraph@NLO}~\cite{Alwall:2014hca} to generate parton-level events. For both signal and backgrounds at the parton level, we apply the following {\it pre-selection cuts}:
\begin{eqnarray}
&p^{j,\ell}_T >  5\,{\rm GeV}, \quad \ |\eta_{j,\ell}|\ < 5, \nonumber\\
&\Delta R_{j\ell} > 0.4, \quad  \Delta R_{\ell\ell} > 0.4,
\label{eq:basic}
\end{eqnarray}
where $p_T$ and $\eta$ are the transverse momentum and pseudorapidity for the corresponding particles, and $\Delta R =\sqrt{\Delta \phi^2+\Delta y^2}$ is
the separation in the azimuthal angle-rapidity ($\phi-y$) plane. Notice that all of the cuts in Eq.~(\ref{eq:basic}) are defined in the lab frame. We then use {\sc Pythia6.4} \cite{Sjostrand:2006za} to implement parton shower and hadronization, and {\sc Delphes3.4} \cite{Ovyn:2009tx} to fast detector simulation according to the LHeC designed parameters~\cite{AbelleiraFernandez:2012cc}.

In order to suppress RB, we apply the following veto criteria~\cite{Han:2018rkz} as {\it basic cuts}:
\begin{itemize}
	\item events are required to contain exactly one hard electron, one hard jet, and ${\slashed E}_T$,
	\item events containing any extra jets with $p_T^j>3$ GeV and $|\eta_j|<2$ or leptons with $p_T^{e,\mu}>5$ GeV or tagged $\tau$ jets are voted.
\end{itemize}

\begin{figure}[!htbp]
\begin{center}
\includegraphics[width=0.45\columnwidth]{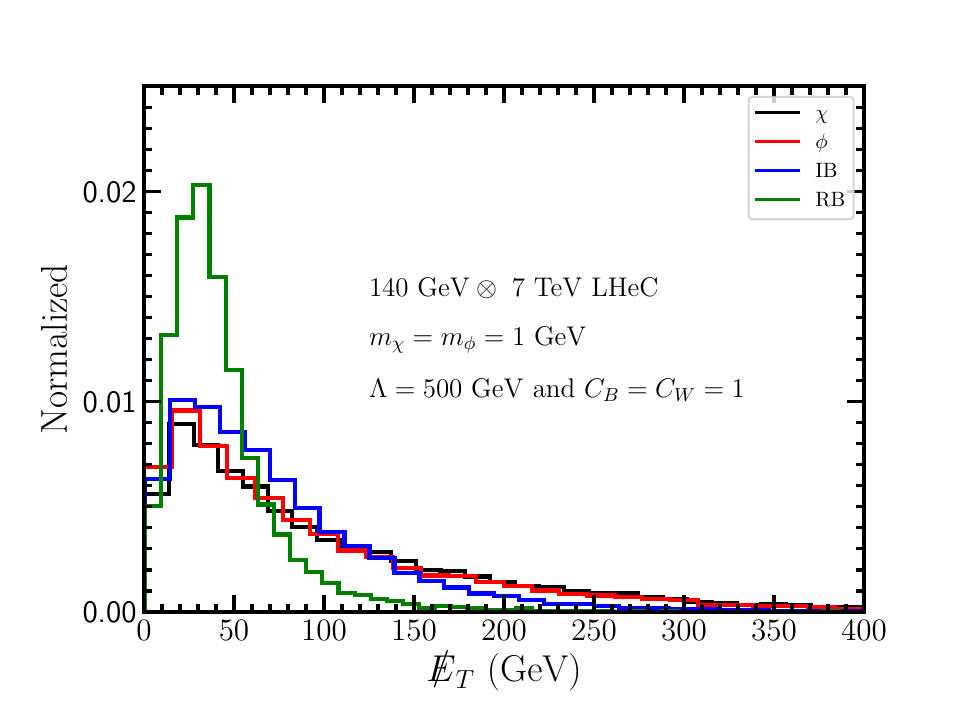}
~~
\includegraphics[width=0.45\columnwidth]{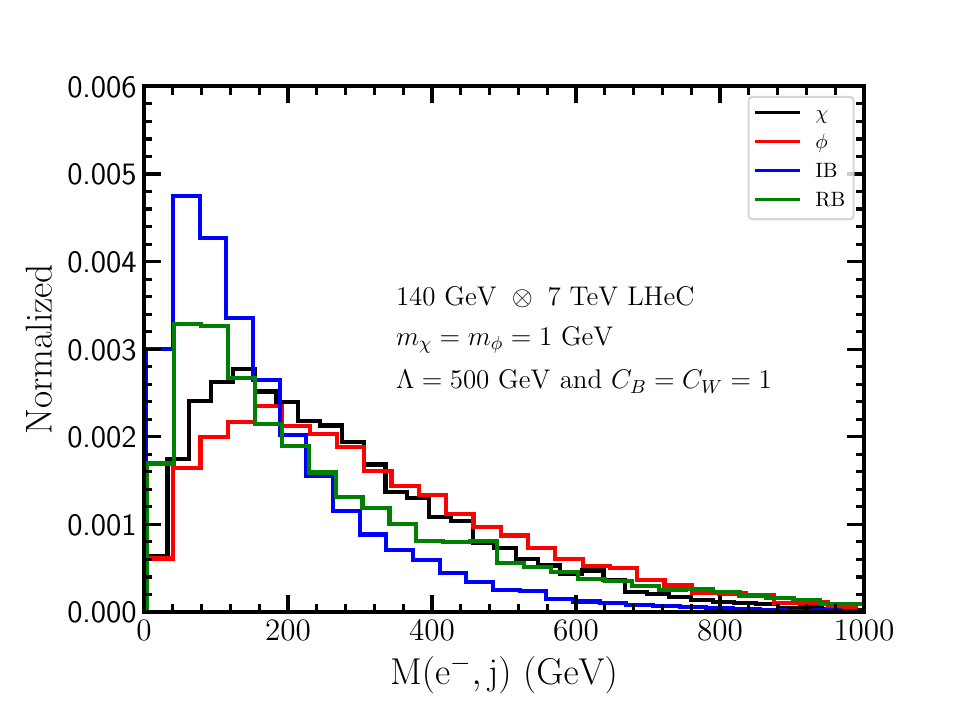}
~~
\includegraphics[width=0.45\columnwidth]{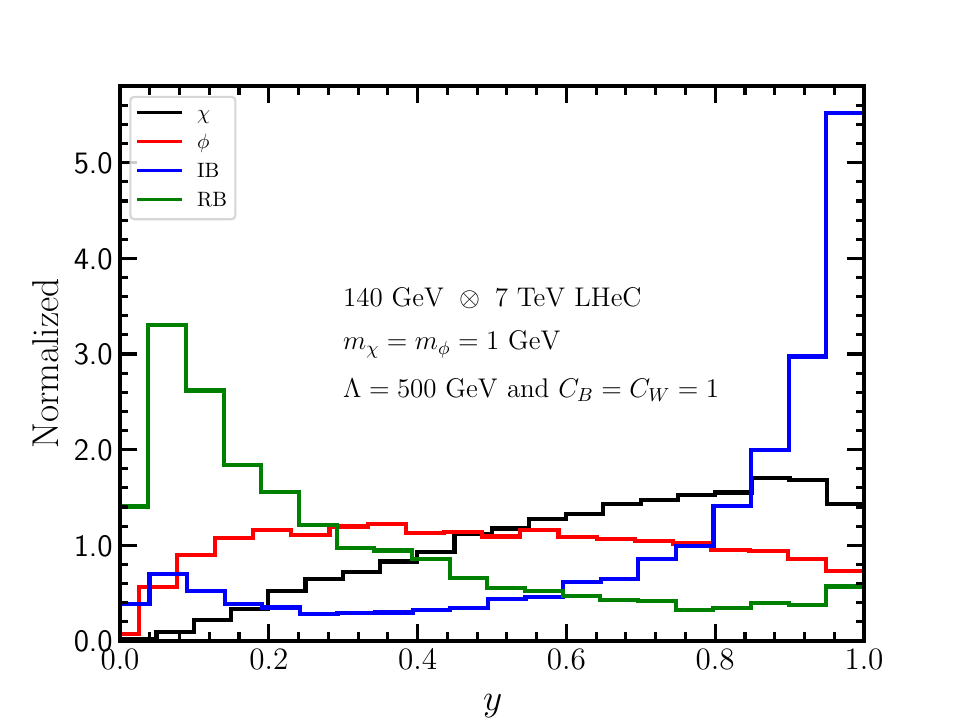}
\end{center}
\caption{The normalized distribution for  the missing transverse energy ${\slashed E}_T$(left), the invariant mass of electron-jet system $M(e^-,j)$ (middel) and the inelasticity variable $y$ (right) with $m_\chi=m_\phi=1$ GeV, $\Lambda=500$ GeV and $C_B=C_W=1$ at the 140 GeV $\otimes$ 7 TeV LHeC. The red and black curves respectively correspond to scalar and fermion DM signals, and SM RB and IB are shown by green and blue lines.
\label{fig:distributionlhec}}
\end{figure}

\begin{figure}[!htbp]
\begin{center}
\includegraphics[width=0.45\columnwidth]{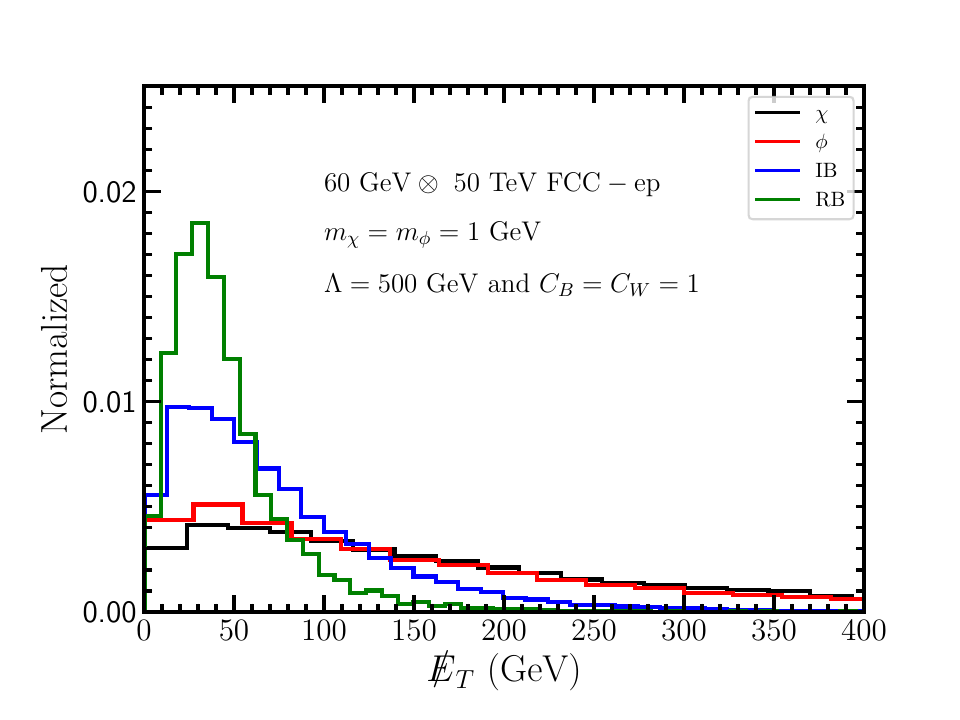}
~~
\includegraphics[width=0.45\columnwidth]{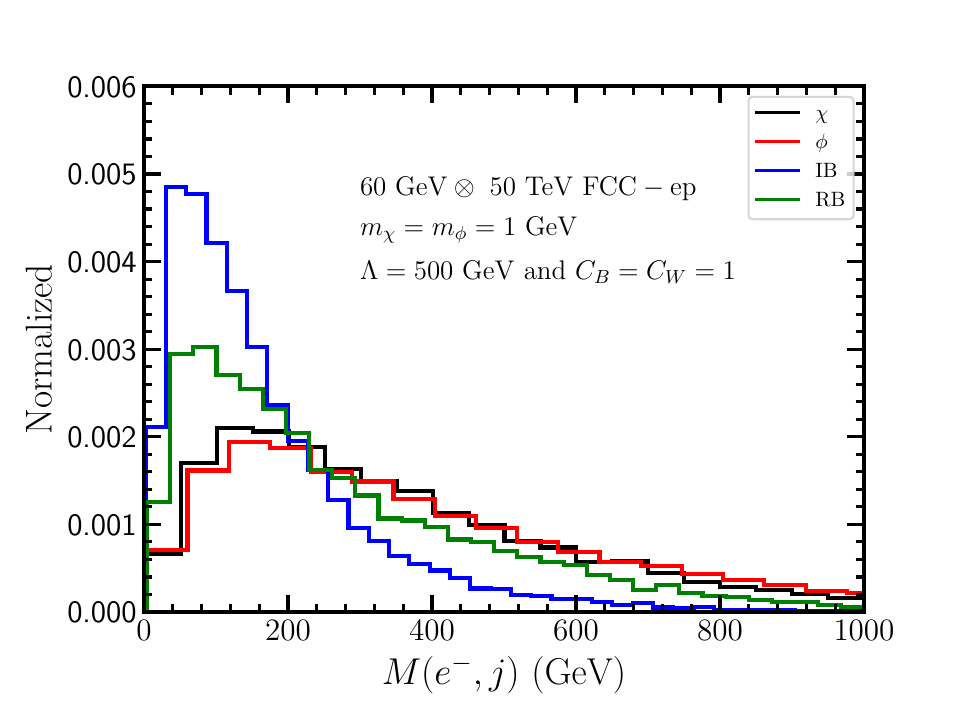}
~~
\includegraphics[width=0.45\columnwidth]{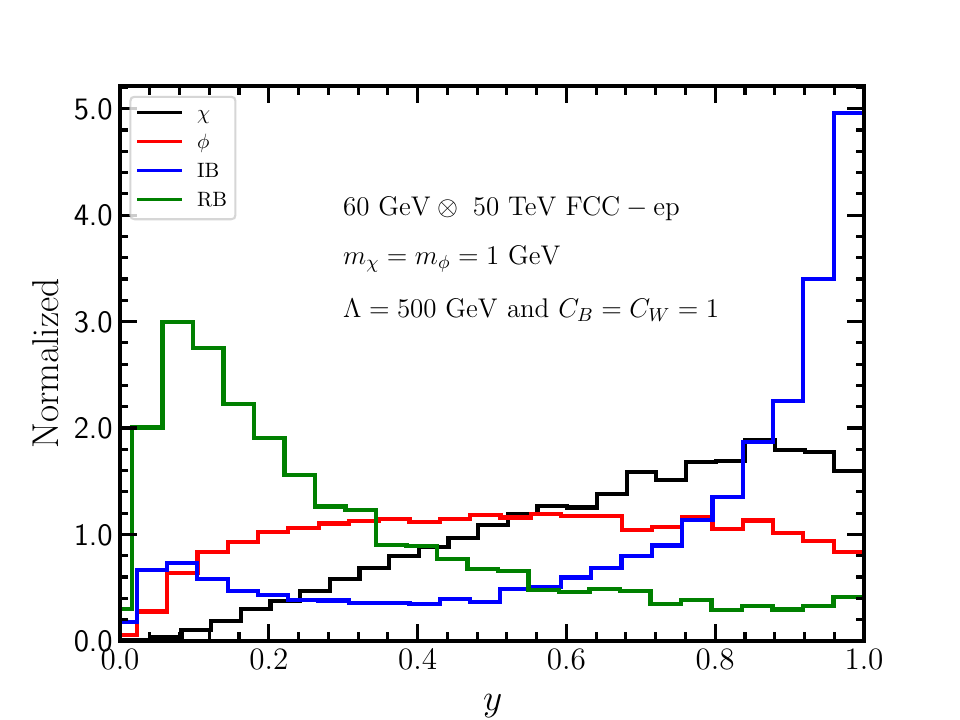}
\end{center}
\caption{Same as Fig. \ref{fig:distributionlhec}, but at the 60 GeV $\otimes$ 50 TeV FCC-ep.
\label{fig:distributionfcc}}
\end{figure}

In Fig.~\ref{fig:distributionlhec} and Fig.~\ref{fig:distributionfcc}, we respectively display the normalized distributions of missing transverse energy ${\slashed E}_T$, the invariant mass of tagged electron and jet system $M(e^-,j)$ and inelasticity variable $y$ for both signal and backgrounds after the {\it basic cuts}, at the 140 GeV $\otimes$ 7 TeV LHeC and 60 GeV $\otimes$ 50 TeV FCC-ep. For signal, we assumed the benchmark values $m_{\chi,~\phi}=1$ GeV, $C_{B,~W}=1$ and $\Lambda=500$ GeV. For inelasticity variable $y$, we follow the definition in Ref.~\cite{Tang:2015uha} which is given as
\begin{eqnarray}
y\equiv\frac{k_P\cdot(k_e-p_e)}{k_P\cdot k_e},
\end{eqnarray}
where $k_{P,~e}$ are the 4-momenta of the initial proton and electron, and $p_e$ is the 4-momenta of the final electron. One can observe the following features:
\begin{itemize}
	\item for ${\slashed E}_T$ distribution, SM RB tends to possess smaller value ${\slashed E}_T <100$ GeV compared with signals, while SM IB has similar with signals,
	\item for $y$ distribution, SM RB and IB respectively tend to distribute at small and large $y$ values, while signals present relatively flat distribution.
\end{itemize}
Such distinct behaviors of signals and SM backgrounds indicate that we can impose the following kinematical cuts for LHeC (FCC-ep) to further extract signals:
\begin{eqnarray}
&{\slashed E}_T > 100\quad(220)\,{\rm GeV}, \nonumber\\
&M(e^-,j) > 100 \quad (200)\,{\rm GeV}, \nonumber\\
&0.3 < y < 0.8 \quad (0.1 < y < 0.9).
\label{eq:kinematical}
\end{eqnarray}

\begin{table*}[hbt]
\begin{tabular}{|c|c|c|c|c|c|c|c|c|}
\hline
Process & Signal ($\phi$) & Signal ($\chi$) & RB & IB  & ${\cal S}_\phi$  & ${\cal S}_\chi$ & ${\cal R}_\phi$  & ${\cal R}_\chi$ \\ \hline
{\it pre-selection cuts}& 7.57 & 10.28&1421.87 & 260.39 &8.26&11.20 &0.0045 &0.0061 \\ \hline
{\it basic cuts} &4.91&6.84 & 316.83&155.21&10.11&14.08& 0.010&0.014\\ \hline
${\slashed E}_T > 100$ GeV & 1.71 &2.62 &18.28 & 38.33&10.16 &15.58&0.030 &0.046\\ \hline
$M(e^-,j) > 100$ GeV & 1.56 &2.33 &15.46 & 29.28&10.46& 15.56& 0.035&0.052\\ \hline
$0.3 < y < 0.8$  & 1.01 &1.54 &5.40 & 9.16&11.80 &18.05&0.070 &0.106\\ \hline
\end{tabular}
\caption{The remaining cross-section (in a unit of fb) of signal and background after corresponding cuts with $m_{\chi,~\phi}=1$ GeV, $C_{B,~W}=1$ and $\Lambda=500$ GeV at 140 GeV $\otimes$ 7 TeV LHeC. The corresponding significance ${\cal S}=N_S/\sqrt{N_B}$ and the signal-to-background ratio ${\cal R}=N_S/N_B$ are obtained for $L=2~{\rm ab}^{-1}$, where $N_S$ and $N_B$ are the number of signal and background events.}
\label{tab:cutflow-lhec}
\end{table*}

\begin{table*}[hbt]
\begin{tabular}{|c|c|c|c|c|c|c|c|c|}
\hline
Process & Signal ($\phi$) &Signal ($\chi$) & RB & IB  & ${\cal S}_\phi$  & ${\cal S}_\chi$ & ${\cal R}_\phi$  & ${\cal R}_\chi$ \\ \hline
{\it pre-selection cuts} & 18.17 & 64.85&1232.28 & 222.80 &21.31&76.03&0.012&0.045\\ \hline
{\it basic cuts} &11.56&42.58&244.95&125.09&26.87&98.98&0.031&0.12\\ \hline
${\slashed E}_T > 220$ GeV  &3.10 &13.44 &0.86& 4.32&60.88 &264.17&0.060&2.60\\	\hline	
$M(e^-,j) > 200$ GeV   & 2.59&10.71 &0.61 & 2.13&70.21& 289.95&0.95&3.92\\ \hline
$0.1 < y < 0.9$ & 2.41&9.48 &0.57 & 1.44&75.94 &298.87&1.19&4.71\\ \hline
\end{tabular}
\caption{Same as Table~\ref{tab:cutflow-lhec}, but at the 60 GeV $\otimes$ 50 TeV FCC-ep.}
\label{tab:cutflow-fcc}
\end{table*}

In Table.~\ref{tab:cutflow-lhec} and Table.~\ref{tab:cutflow-fcc}, we show the cut-flows for the fermion and scalar DM signals and backgrounds at LHeC and FCC-ep, respectively. The corresponding significance ${\cal S}=N_S/\sqrt{N_B}$ and signal-to-background ratio ${\cal R}=N_S/N_B$ with the total integrated luminosity $L = 2~{\rm ab}^{-1}$ are also shown, where $N_{S,~B}=\sigma_{S,~B}\cdot L$, are respectively the survived event number of signals and backgrounds. One can see that both ${\cal S}$ and ${\cal R}$ get increasing with each cut imposed. After the {\it pre-selection cuts}, RB contribution is still huge, while the {\it basic cuts} suppress RB by a factor of 4.5 (5.0) at the LHeC (FCC-ep). Moreover, the RB is dramatically reduced by the ${\slashed E}_T$ cut, which only survived about $1.26\%$ (0.07\%) at the LHeC (FCC-ep).

At the LHeC, after all cuts, the fermion (scalar) DM signal can survive $13.2\%$ ($14.9\%$), while RB (IB) only survived $0.37\%$ ($3.5\%$). At the FCC-ep, the signal for fermion (scalar) DM survived $13.3\%$ ($14.6\%$), and RB (IB) survived $0.047\%$ ($0.65\%$) at the end. It is worth noting that the remaining cross-section of background at the FCC-ep are only $14\%$ of ones that at the LHeC, while signal for fermion (scalar) DM production at the FCC-ep could reach  $6$ ($1.5$) times than that at the LHeC, which implies FCC-ep has considerable improvement on significance compared with LHeC.

\section{Sensitivity at the CEPC in three different modes}\label{sec:cepc}

\begin{figure}[htbp]
	\vspace{0.2cm}
	\centering
	\includegraphics[scale=0.8]{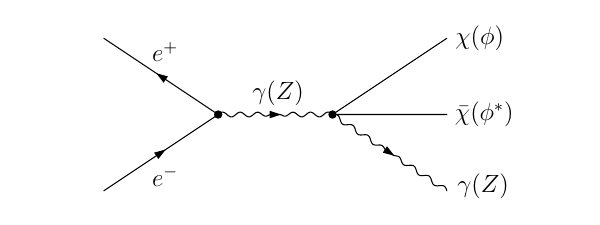}
	\caption{Feynman diagrams for the processes $e^+ e^-  \to \phi\phi^*(\bar \chi \chi) + Z$ or $e^+ e^-  \to \phi\phi^*(\bar \chi \chi) + \gamma$.}	
	\label{fig:process-cepc}
\end{figure}

Due to the effective DM-diboson operators in Eqs.~(\ref{eq:ppvv}) and ~(\ref{eq:ccvv}), the WIMP can be
produced at $e^+e^-$ colliders via the processes $e^+ e^- \to V^{\prime*} \to \phi\phi^*(\bar \chi \chi) + V\ (V=\gamma,Z)$, which give rise to
mono-photon or mono-$Z$ signatures. The corresponding Feynman diagrams
are displayed in Fig. \ref{fig:process-cepc}.
In this section, we will focus on the mono-photon signature at future projected CEPC \cite{CEPCStudyGroup:2018rmc, CEPCStudyGroup:2018ghi} with three different running modes, including
the Higgs factory mode for seven years running
at $\sqrt{s}=240\ \mathrm{GeV}$ with a total luminosity of $\sim 5.6\ \mathrm{ab}^{-1}$, the $Z$ factory mode
for two years running at $\sqrt{s}=91.2\ \mathrm{GeV}$ with a total luminosity of $\sim 16\ \mathrm{ab}^{-1}$, and the $W^+W^-$ threshold scan mode for one year running at
$\sqrt{s} \sim$ $158-172\ \mathrm{GeV}$ with a total luminosity of $\sim 2.6\ \mathrm{ab}^{-1}$.~\footnote{We take $\sqrt s= 160$ GeV for the $W^+W^-$ threshold scan mode throughout our analysis.}

For the monophoton signature at CEPC, an irreducible background arises from the neutrino production $e^+e^-\to\nu_l\bar\nu_l\gamma$ processes, where $\nu_l=\nu_e,\nu_\mu,\nu_\tau$. Due to the Breit-Wigner distribution of $Z$ boson in the irreducible BG, there is a resonance in the final photon
energy spectrum, which is located at
\be
E_\gamma^Z=\frac{s-M_Z^2}{2\sqrt{s}}
\ee
with a full-width-at-half-maximum as
$\Gamma_\gamma^Z=M_Z\Gamma_Z/\sqrt{s}$.
In order to suppress the irreducible background contribution,
we will veto the events within $5\Gamma_\gamma^Z$ at the $Z$ resonance in
the monophoton energy spectrum \cite{Liu:2019ogn}.
The vetoing cut can be written as
\be
|E_\gamma-E_\gamma^Z|<5\Gamma_\gamma^Z.
\ee

The major reducible SM backgrounds come from the $e^+e^-\to\gamma+{\slashed X}$ processes,
where only one photon is visible in the final state, and ${\slashed X}$ denotes the other
undetectable particle(s) due to the limited detection capability of the detectors.
In our analysis, the  parameters for the EMC coverage: $|\cos\theta| <0.99$
and $E>0.1$ GeV, are adopted following CEPC CDR \cite{CEPCStudyGroup:2018ghi}.
The processes $e^+e^-\to f\bar f \gamma$ and $e^+e^-\to\gamma\gamma\gamma$ provide
dominate contributions to the reducible background when final $f\bar f$ and $\gamma\gamma$
are emitted with $|\cos\theta| >0.99$.
Due to the momentum conservation in the transverse direction and energy conservation, the maximum photon energy as a function of its polar angle can be obtained as  \cite{Liu:2019ogn}
\be
E_\gamma^m(\theta_\gamma) =
\sqrt{s}\left(\frac{\sin\theta_\gamma}{\sin\theta_b}\right)^{-1},
\label{eq:bBG-besiii}
\ee
where the polar angle $\theta_b$ corresponds to the boundary of the EMC with
$|\cos\theta| <0.99$.
In order to suppress the monophoton events arising from the reducible background,
we adopt the detector cut $E_\gamma>E_\gamma^m(\theta_\gamma)$ on the final state
photon in our analysis.

\section{Update limits from the LHC mono-$j$ and mono-$\gamma$ searches}\label{sec:LHC}
\begin{figure}[htbp]
	\vspace{0.2cm}
	\centering
	\includegraphics[scale=0.7]{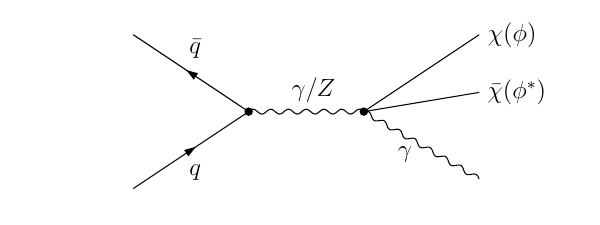}
	\includegraphics[scale=0.7]{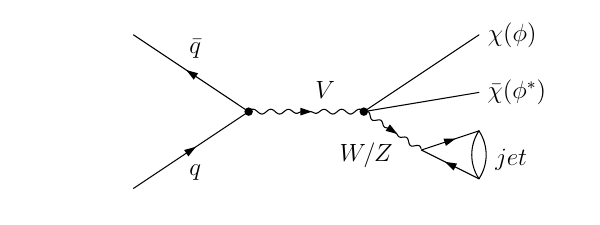}
	\caption{Representative Feynman diagrams for a mono-photon (Left) and a mono-jet signature  (Right)
		arising from DM-diboson operators.}	
	\label{fig:tree-fd}
\end{figure}

\begin{figure}[htbp]
	\vspace{0.2cm}
	\centering
	\includegraphics[scale=0.7]{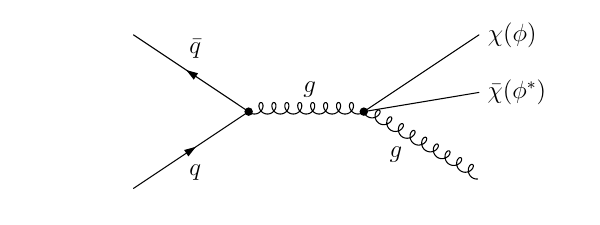}
	\includegraphics[scale=0.7]{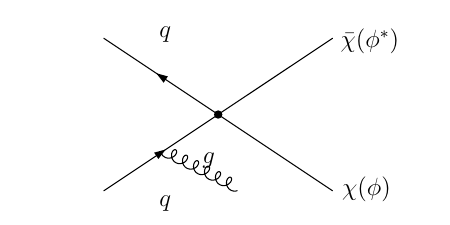}
	\includegraphics[scale=0.7]{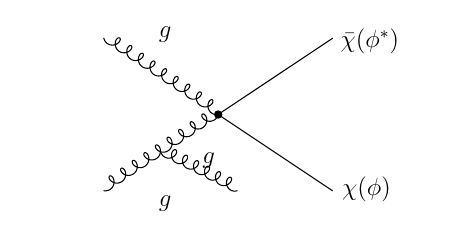}
	\includegraphics[scale=0.7]{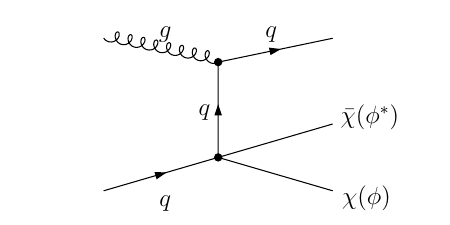}
	\includegraphics[scale=0.7]{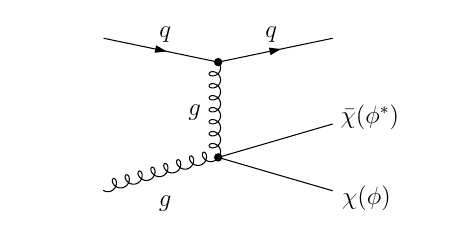}
	\caption{Representative Feynman diagrams for a mono-jet signature
		arising from loop-induced effective DM-quarks and DM-gluons operators.}	
	\label{fig:loop-fd}
\end{figure}

Due to the DM-diboson operators, the WIMP can be
produced via the processes $q q^\prime \to V^{\prime*} \to \phi\phi^*(\bar \chi \chi) + V\ (V=\gamma,Z,W)$,
which can exhibit a mono-photon (${\slashed E_T+\gamma)}$ or
a mono-jet (${\slashed E_T+W/Z\ (\to {\rm hardons} )}$) signature, respectively.
The representative Feynman diagrams are shown in Fig. \ref{fig:tree-fd}.
Besides, the loop-induced effective DM-quarks (${\cal O}_q$) and DM-gluons (${\cal O}_G$) operators, could also result in the mono-jet signature via
$q \bar q \to \bar\chi \chi +g$, $gg \to \bar\chi \chi +g$, and
$q g \to \bar\chi \chi +q$ processes.
The representative Feynman diagrams are shown in Fig. \ref{fig:loop-fd}.

In this section, we consider the sensitivities from LHC mono-$j$ and mono-$\gamma$ searches.
To simulate the signal from dark matter, the parton-level events are generated by {\sc MadGraph@NLO}~\cite{Alwall:2014hca}, and the following parton shower and hadronization are dealt with
{\sc Pythia6.4} \cite{Sjostrand:2006za}. We use {\sc Delphes3.4} \cite{Ovyn:2009tx} to carry out a
fast detector simulation for the ATLAS or CMS detector with the corresponding parameter setup.
We follow the procedure of the mono-photon \cite{Aad:2020arf} and mono-jet \cite{Aad:2021egl} latest analysis by the ATLAS Collaboration with an integrated luminosity of 139 fb$^{-1}$ at a center-of-mass energy of 13 TeV.

In the case of ${\slashed E_T+\gamma}$ search channel, events in our analysis are required to have a leading $\gamma$ with
$
E_{{T}}^{\gamma}>150 \mathrm{GeV},\ |\eta^\gamma|<1.37 \text { or } 1.52<|\eta^\gamma|<2.37,\ \Delta \phi\left(\gamma,\  \boldsymbol{E}_{{T}}^{\text {miss }}\right)>0.4,
$
0 or 1 jet with
$
p_{{T}}>30 \mathrm{GeV},\ |\eta|<4.5 \text { and } \Delta \phi\left(\mathrm{jet},\ \boldsymbol{E}_{{T}}^{\mathrm{miss}}\right)>0.4
$,
and none leptons. Besides, the ATLAS Collaboration performs the measurement in seven different signal regions with a varying cut on the missing transverse momentum ${E}_{{T}}^{\mathrm{miss}}$.
We find that the strongest bounds on the parameter space in our case are given with the most severe cut of ${E}_{{T}}^{\mathrm{miss}}> 375 $ GeV. The corresponding 95\% C.L. limit on the fiducial cross-section reads $\sigma_{\rm fid}< $ 0.53 fb.

For mono-jet search, events having identified muons, electrons, photons or $\tau$-leptons in the final state are vetoed. Selected events have a leading jet with $p_T>150$ GeV and $|\eta|<2.4$, and up to three additional jets with $p_T>30$ GeV and $|\eta|<2.8$. Separation in the azimuthal angle
between the missing transverse momentum direction and each selected jet $\Delta \phi\left(\mathrm{jet},\ \boldsymbol{E}_{{T}}^{\mathrm{miss}}\right)>0.4 (0.6)$ is required for events with ${E}_{\mathrm{T}}^{\mathrm{miss}}>250$ GeV ($200 \mathrm{GeV} < {E}_{{T}}^{\mathrm{miss}}< 250$ GeV). It turns out that the stronger restriction of ${E}_{{T}}^{\mathrm{miss}} > 1200$  GeV provides the best limits. At 95\% C.L. the bound on the corresponding fiducial cross section
is given by $\sigma_{\rm fid}< $ 0.3 fb.

In Tab. \ref{tab:ratio}, we present the ratio of the number of events for mono-jet signature at LHC from the process ${pp\to \chi\bar\chi+W/Z\ (\to {\rm hardons} )}$ via dimension-7 DM-diboson operators with the Feynman diagrams shown in Fig. \ref{fig:tree-fd} to one from the process ${pp\to \chi\bar\chi+ q/g}$ via loop-induced  DM-quarks and DM-gluons operators with the Feynman diagrams shown in Fig. \ref{fig:loop-fd} with $\Lambda$=1000 GeV. There we consider three cases, i.e., $C_{B}=C_{W}=500$, $C_{B}=0, C_{W}=500$, and $C_{B}=500,~C_{W}=0$. It can be found that the ratios in the three cases all increase with the increment of the DM mass. Besides, to the mono-jet signature at LHC the contributions from tree-level DM-diboson operators are about two orders of magnitudes over from loop-induced  DM-quarks and DM-gluons operators, especially about three orders in the case of $C_{B}=1,~C_{W}=0$, thus we ignore the contribution from DM-quarks and DM-gluons operators in our following analysis.

\begin{table}[h]
	\begin{tabular}{|c|c|c|c|c|c|}
		\hline
		\hline
		$m_\chi$ & 1 &   300&   800& 2000 & 3000 \\
		\hline
		$C_B=500,C_W=0$ &638.22 & 699.35  & 752.11 & 1435.32 & 2733.95\\
		\hline
		$C_B=0,C_W=500$ &95.00 & 99.55  & 104.29 & 181.55 & 302.49\\
		\hline
		$C_B=500,C_W=500$ &79.20 & 80.33  & 91.64 & 158.94 & 271.41\\
		\hline\hline
	\end{tabular}
	\caption{The ratio of the number of events for the mono-jet signature at LHC from the process ${pp\to \chi\bar\chi+W/Z\ (\to {\rm hardons} )}$ via DM-diboson operators with the Feynman diagrams shown in Fig. \ref{fig:tree-fd} to ${pp\to \chi\bar\chi+ q/g}$ via loop-induced  DM-quarks and DM-gluons operators with the Feynman diagrams shown in Fig. \ref{fig:loop-fd} with $\Lambda$=1000 GeV.}
	\label{tab:ratio}
\end{table}
\section{Relic abundance, Direct and Indirect Constraints}\label{sec:direct}

In this section, we discuss the bounds on $\Lambda$ from current direct and indirect search experiments.
The effective operators ${\cal O}_B$ and ${\cal O}_W$
lead to the non-relativistic cross-sections of DM annihilation $\chi\bar\chi,\ \phi\phi^*\to\gamma\gamma/\gamma Z/ZZ/WW$ as
\be
\langle\sigma_k v\rangle = a_k + b_k v^2+ {\cal O}(v^4)
\ee
with $k=B,W$ annihilation channels. The coefficients of $a$ and $b$ are \cite{Chen:2013gya}
\begin{small}
\begin{align}
\chi\left\{\begin{aligned}
a_B &=a_W=0,\\
b_B &=\frac{\left|C_{B}(\Lambda)\right|^{2}}{\pi} \frac{m_{\chi}^{4}}{\Lambda^{6}}\Big[c_{w}^{4}+\frac{c_{w}^{2} s_{w}^{2}}{8} \beta_{\gamma Z}^{2}\left(x_{Z}-4\right)^{2}\Theta(M_Z<2m_\chi) +\frac{s_{w}^{4}}{8} \beta_{Z Z}\left(3 x_{Z}^{2}-8 x_{Z}+8\right)\Theta(M_Z<m_\chi)\Big], \\
b_W &=\frac{\left|C_{W}(\Lambda)\right|^{2}}{\pi} \frac{m_{\chi}^{4}}{\Lambda^{6}}\Big[s_{w}^{4}+\frac{c_{w}^{2} s_{w}^{2}}{8} \beta_{\gamma Z}^{2}\left(x_{Z}-4\right)^{2}\Theta(M_Z<2m_\chi) +\frac{c_{w}^{4}}{8} \beta_{Z Z}\left(3 x_{Z}^{2}-8 x_{Z}+8\right)\Theta(M_Z<m_\chi)\\
    &+\frac{1}{4} \beta_{W W}\left(3 x_{W}^{2}-8 x_{W}+8\right)\Theta(M_W<m_\chi)\Big],
\end{aligned}\right.
\end{align}
\begin{align}
\phi\left\{\begin{aligned}
a_B &= \frac{\left|C_{B}(\Lambda)\right|^{2}}{\pi} \frac{2m_{\phi}^{2}}{\Lambda^{4}}\Big[c_{w}^{4} +\frac{c_{w}^{2} s_{w}^{2}}{8} \beta_{\gamma Z}^{2}\left(x_{Z}-4\right)^{2}\Theta(M_Z<2m_\phi)+\frac{s_{w}^{4}}{8} \beta_{Z Z}\left(3 x_{Z}^{2}-8 x_{Z}+8\right)\Theta(M_Z<m_\phi)\Big],\\
a_W &= \frac{\left|C_{W}(\Lambda)\right|^{2}}{\pi} \frac{2m_{\phi}^{2}}{\Lambda^{4}}\Big[s_{w}^{4} +\frac{c_{w}^{2} s_{w}^{2}}{8} \beta_{\gamma Z}^{2}\left(x_{Z}-4\right)^{2}\Theta(M_Z<2m_\phi)+\frac{c_{w}^{4}}{8} \beta_{Z Z}\left(3 x_{Z}^{2}-8 x_{Z}+8\right)\Theta(M_Z<m_\phi) \\
&+\frac{1}{4} \beta_{W W}\left(3 x_{W}^{2}-8 x_{W}+8\right)\Theta(M_W<m_\phi)\Big], \\
b_{B} &=\frac{\left|C_{B}(\Lambda)\right|^{2}}{\pi} \frac{2m_{\phi}^{2}}{\Lambda^{4}}\Big[c_{w}^{4}/2 +\frac{c_{w}^{2} s_{w}^{2}}{4} \beta_{\gamma Z}^{2}\left(x_{Z}-4\right)\Theta(M_Z<2m_\phi)+s_{w}^{4} \beta_{Z Z}\left(1/2-x_Z\right)\Theta(M_Z<m_\phi)\Big], \\
b_{W} &=\frac{\left|C_{W}(\Lambda)\right|^{2}}{\pi} \frac{2m_{\phi}^{2}}{\Lambda^{4}}\Big[s_{w}^{4}/2 +\frac{c_{w}^{2} s_{w}^{2}}{4} \beta_{\gamma Z}^{2}\left(x_{Z}-4\right)\Theta(M_Z<2m_\phi)+c_{w}^{4} \beta_{Z Z}\left(1/2-x_Z\right)\Theta(M_Z<m_\phi)\\
&+\beta_{W W}\left(1-2x_Z\right)\Theta(M_W<m_\phi)\Big].
\end{aligned}\right.
\end{align}
\end{small}
In above equations, we have defined the phase space factor $\beta_{i j}=\sqrt{1-\left(m_{i}+m_{j}\right)^2/\left(4 m_{\rm DM}^2\right)}$ and $x_i=m_i^2/m_{\rm DM}^2$ with ${\rm DM}=\chi,\phi$.
Using the results for $a_{B,W}$ and $b_{B,W}$, the relic density can then be calculated with
\begin{equation}
\Omega_{\rm DM} h^{2}=\frac{1.09 \times 10^{9} x_{F} \mathrm{GeV}^{-1}}{M_{P l} \sqrt{g_{*}\left(x_{F}\right)}\left(a+3 b / x_{F}\right)},
\end{equation}
where $a=a_B+a_W$ and $b=b_B+b_W$.
$x_F=m_{\rm DM}/T_F$, being the ratio of the mass of DM $m_{\rm DM}$ and the early-universe freeze-out temperature $T_F$, can be obtained by solving
\begin{equation}
    x_F=\ln\left[c(c+2)\sqrt{\frac{45}{8}}\frac{m_{\rm DM}M_{Pl}(a+6b/x_F)}{2\pi^3\sqrt{g_*(x_F)}}\right].
\end{equation}


The loop-induced effective DM-quarks (${\cal O}_q$) and DM-gluons (${\cal O}_G$) operators, could result in the interaction between DM and nucleon. The spin-independent (SI) cross section for elastic scalar and Dirac WIMP scattering on a nucleon has the form
\bea
\sigma_{\rm SI}^{\mathrm{\phi}} &\simeq& \frac{\mu_N^{2} m_{N}^{2}}{\pi m_\phi^2 \Lambda^{4}}\left|\frac{\alpha Z^{2}}{A} f_{A}^{N} C_{AA}\left(\mu_{\rm had}\right)+\frac{Z}{A} f_{p}\left(\mu_{\rm had}\right)+\frac{A-Z}{A} f_{n}\left(\mu_{\rm had}\right)\right|^{2}\,, \\
\sigma_{\rm SI}^{\mathrm{\chi}} &\simeq& \frac{\mu_N^{2} m_{N}^{2}}{\pi \Lambda^{6}}\left|\frac{\alpha Z^{2}}{A} f_{A}^{N} C_{AA}\left(\mu_{\rm had}\right)+\frac{Z}{A} f_{p}\left(\mu_{\rm had}\right)+\frac{A-Z}{A} f_{n}\left(\mu_{\rm had}\right)\right|^{2}\,.
\eea
In above equation, $m_N\simeq 0.939$ GeV is the average nucleon mass, and $\mu_N=m_{\phi,\chi}m_N/(m_{\phi,\chi}+m_N)$ is the reduced mass of the DM-nucleon system. The form factor $f_{A}^{N}$ describes the Rayleigh scattering of two photons on the entire nucleus. At zero-momentum transfer limit, $f_{A}^{N}\simeq0.08$ for xenon target. The form factor $f_{N=p,n}$ describing the couplings between the DM and nucleon reads~\cite{Weiner:2012cb,Frandsen:2012db}
\begin{equation}
f_{N}\left(\mu_{\rm had}\right)=\sum_{q=u, d, s} f_{q}^{N} C_{q}\left(\mu_{\rm had}\right)-\frac{8 \pi}{9} f_{G}^{N} C_{G}\left(\mu_{\rm had}\right),
\end{equation}
where from factors $f_{q}^{N}$ describing  the scalar couplings between quarks and nucleon are given by~\cite{Belanger:2013oya}

\begin{equation}
f_{d}^{p}=0.0191, \ f_{u}^{p}=0.0153, \ f_{d}^{n}=0.0273,\ f_{u}^{n}=0.0110,\  f_{s}^{p}=f_{s}^{n}=0.0447
\end{equation}
and
\begin{equation}
f_{G}^{N}=1-\sum_{q=u, d, s} f_{q}^{N}.
\end{equation}

\section{Results}\label{sec:results}

\begin{figure}[!htbp]
\begin{center}
\includegraphics[width=\linewidth]{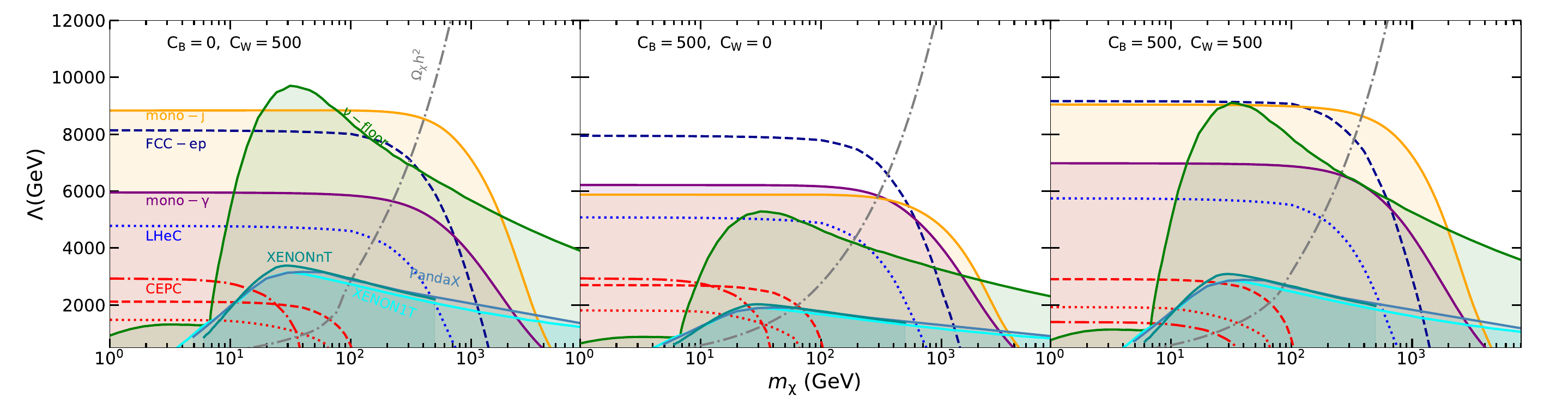}
\end{center}
\caption{
Constraints in the $m_\chi-\Lambda$ plane for the fermion DM with $C_B^{6(7)}=0,C_W^{6(7)}=500$ (left panel) , $C_B^{6(7)}=500,C_W^{6(7)}=0$ (middle panel) and $C_B^{6(7)}=500,C_W^{6(7)}=500$ (right panel).
purple and orange solid lines denote the exclusion limits from the mono-photon \cite{Aad:2020arf} and mono-jet \cite{Aad:2021egl} searches at the 95\% C.L. at the 13 TeV LHC. For the direct searches, recent bounds from
XENON1T \cite{XENON:2017vdw,XENON:2018voc}, XENONnT \cite{XENON:2023sxq} and PandaX-4T \cite{PandaX-4T:2021bab} are shown as cyan, dark cyan and steel blue solid lines, respectively. For illustration, the contours of the relic abundance  $\Omega_{\rm DM} h^{2}=0.1186$ are also plotted with gray dash-dotted curves. The neutrino floor is shown in green-shaded regions. The expected 95\% C.L. bounds at the future LHeC (blue dotted lines), FCC-ep (dark blue dashed lines),
and CEPC (red dotted lines for $Z$ factory mode, red dash-dotted lines for $W^+W^-$ threshold scan mode, red dashed lines for Higgs factory mode) are shown.
\label{fig:limit_chi}}
\end{figure}

\begin{figure}[!htbp]
\begin{center}
\includegraphics[width=\linewidth]{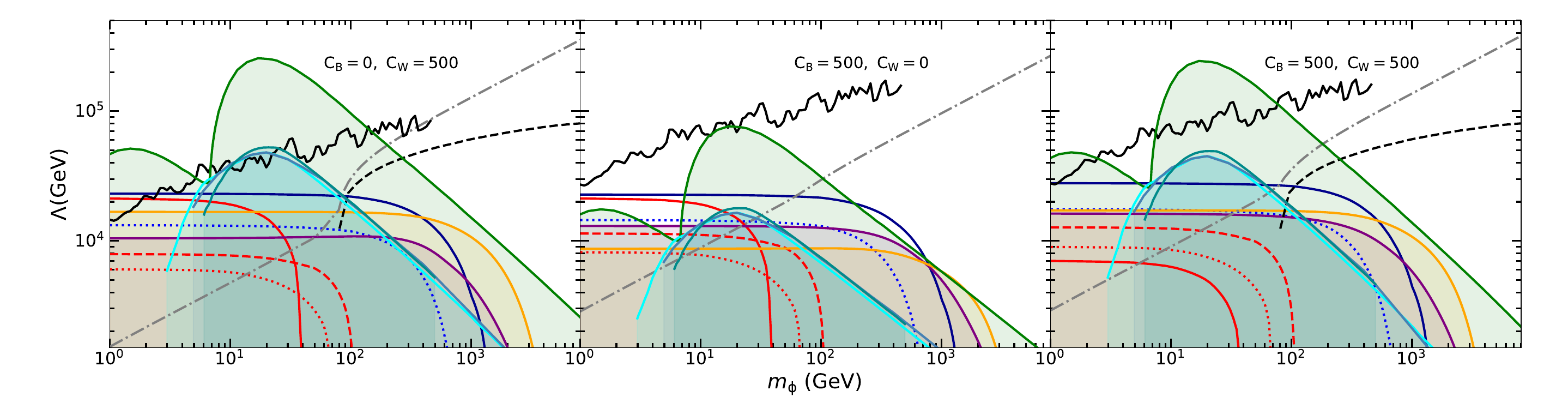}
\end{center}
\caption{Same as Fig. \ref{fig:limit_chi}, but in the $m_\phi-\Lambda$ plane for the fermion DM. The constraints from the indirect search of DM through the cosmic gamma-ray (black solid line) and $WW$ (black dashed line) observations by the Fermi-LAT collaboration~\cite{Fermi-LAT:2015att} are also shown.
\label{fig:limit_phi}}
\end{figure}

We apply simple $\chi^2$ analysis to derive the lower bounds on $\Lambda$ for effective operators in Eq.~(\ref{eq:lag})  at future LHeC, FCC-ep, and CEPC at $95\%$ C.L., by requiring $\chi^2=N_S^2/N_B=3.84$~\cite{Zyla:2020zbs} with the specified luminosities.
For LHeC and FCC-ep, the total integrated luminosities are assumed as $L = 2~{\rm ab}^{-1}$.
The limits of $\Lambda$ as a function of the mass of Dirac fermion and scalar DM
are displayed in Figs.~\ref{fig:limit_chi} and \ref{fig:limit_phi}, respectively.
We also present the exclusion limits derived from the mono-photon \cite{Aad:2020arf} search and mono-jet ~\cite{Aad:2021egl} search at the LHC and from the direct DM experiments XENON1T~\cite{XENON:2017vdw,XENON:2018voc}, XENONnT \cite{XENON:2023sxq} and PandaX-4T~\cite{PandaX-4T:2021bab}. For illustration, we plot the contours of the relic abundance $\Omega_{\rm DM} h^{2}=0.1186\pm0.0020$ \cite{Planck:2018vyg}. The neutrino floor is also shown, which represents the WIMP-discovery limit obtained with an assuming exposure of 1000 $^8B$ neutrinos detected on a Xenon target \cite{Ruppin:2014bra}. For the scalar DM, we also show the constraints from the DM indirect searches through the gamma-ray and $WW$ observations by Fermi-LAT collaboration~\cite{Fermi-LAT:2015att}. Here we consider three typical cases, i.e., $C_{B}=0,~C_{W}=500$, $C_{B}=500,~C_{W}=0$, and $C_{B}=C_{W}=500$.

For fermion DM, the bounds from CEPC with all the three running modes can touch the neutrino floor with the DM mass less than a few GeV, while the $e^+e^-$ collider CEPC is not competitive with other $ep$ and $pp$ colliders. At LHC, the sensitivity from mono-jet search is better with $C_{B}=0, C_{W}=500$, and $C_{B}=C_{W}=500$, but is slightly worse with  $C_{B}=500,~C_{W}=0$ than mono-photon search. Besides, the LHC mono-jet search can give the best constraint with $C_{B}=0, C_{W}=500$ compared with other collider experiments. From the figures, it is further validated that FCC-ep has great advantages to constrain DM-gauge boson effective interactions compared with LHeC as mentioned before. In the cases of $C_{B}=500,~C_{W}=0$ and $C_{B}=C_{W}=500$, FCC-ep can touch the region unconstrained by other experiments when the DM mass in less than several hundred GeV, and push the limits up to about 8000 GeV and 9000 GeV respectively.

For scalar DM, in the case of $C_{B}=0, C_{W}=500$ all the collider searches cannot escape the neutrino floor, while the indirect searches of DM  by the Fermi-LAT collaboration provide the leading constraints and can touch the neutrino floor in the mass region of $5\ {\rm GeV} \lesssim m_\phi \lesssim 7$ GeV and $m_\phi \gtrsim 130 $ GeV with gamma-ray observation, and $m_\phi\gtrsim280 $ GeV with $WW$ observation. In the case of $C_{B}=500,~C_{W}=500$, except FCC-ep can reach the neutrino floor in the mass region of $5\ {\rm GeV} \lesssim m_\phi \lesssim 6$ GeV, other collider searches cannot escape the neutrino floor. Similar to the case of $C_{B}=0, C_{W}=500$, DM searches by the Fermi-LAT collaboration can touch the neutrino floor for $3\ {\rm GeV} \lesssim m_\phi \lesssim 7$ GeV and $m_\phi \gtrsim 80$  GeV with cosmic gamma-ray observation, and $m_\phi > 250 $ GeV with $WW$  observation. In the case of $C_{B}=500,~C_{W}=0$, the cosmic gamma-ray observation by the Fermi-LAT collaboration almost lies above the neutrino floor in the plotted region, and $WW$  observation cannot provide any constraint since $\phi\bar\phi W^+W^-$ coupling vanishes. For the collider searches, the future FCC-ep provides the most stringent restrictions relative to the other colliders for all the considered three cases with $m_\phi$ less than a few hundred GeV. Besides, in the case of $C_{B}=500,~C_{W}=0$, FCC-ep and CEPC in the $Z$ factory mode can probe a light DM ( $ m_\phi \lesssim 7$ GeV ) touching the neutrino floor.

\section{Summary and conclusion}\label{sec:summary}
In this work, we focus on the constraints on the dimension-6 (7) effective operators between scalar (Dirac fermion) DM and SM gauge bosons at future $ep$ colliders LHeC and FCC-ep via ${\slashed E}_T+e^-j$ signature for the first time. We find that SM irreducible background and reducible background can be effectively suppressed by imposing appropriate kinematic cuts.
We also consider the sensitivity at the future $e^+e^-$ collider CEPC in three different modes with mono-photon signature and update the limits from the LHC with current mono-$j$ and mono-$\gamma$ searches. Besides the contribution from ${\slashed E_T+W/Z\ (\to {\rm hardons} )}$ channel due to the DM-diboson operators for mono-$j$ at LHC, we also investigate the contribution from the loop induced effective DM-quarks and DM-glouns operators. We find that the contribution from DM-quarks and DM-gluons operators can be ignored.
We present the constraints on effective energy scale $\Lambda$ as a function of DM mass with three typical cases of Wilson coefficients: $C_{B}^{6(7)}=C_{W}^{6(7)}=500$, $C_{B}^{6(7)}=0, C_{W}^{6(7)}=500$, and $C_{B}^{6(7)}=500, C_{W}^{6(7)}=0$. The FCC-ep presents better sensitivity than LHeC in all cases for scalar and Dirac fermion DM. It can be found that the collider searches can probe light DM touching the neutrino floor.

\acknowledgments
RD is supported in part by the National Key R\&D Programme of China (2021YFC2203100). YZ is supported in part by the National Natural Science Foundation of China (Grant no. 11805001) and the Fundamental Research Funds for the Central Universities (Grant no.JZ2023HGTB0222).


\begin{thebibliography}{1}
	
\bibitem{Aghanim:2018eyx}
N.~Aghanim \textit{et al.} [Planck],
Astron. Astrophys. \textbf{641} (2020), A6
[erratum: Astron. Astrophys. \textbf{652} (2021), C4]
doi:10.1051/0004-6361/201833910
[arXiv:1807.06209 [astro-ph.CO]].

\bibitem{Lee:1977ua}
B.~W.~Lee and S.~Weinberg,
Phys. Rev. Lett. \textbf{39} (1977), 165-168
doi:10.1103/PhysRevLett.39.165

\bibitem{Jungman:1995df}
G.~Jungman, M.~Kamionkowski and K.~Griest,
Phys. Rept. \textbf{267} (1996), 195-373
doi:10.1016/0370-1573(95)00058-5
[arXiv:hep-ph/9506380 [hep-ph]].

\bibitem{Bertone:2004pz}
G.~Bertone, D.~Hooper and J.~Silk,
Phys. Rept. \textbf{405} (2005), 279-390
doi:10.1016/j.physrep.2004.08.031
[arXiv:hep-ph/0404175 [hep-ph]].

\bibitem{Goodman:2010ku}
J.~Goodman, M.~Ibe, A.~Rajaraman, W.~Shepherd, T.~M.~P.~Tait and H.~B.~Yu,
Phys. Rev. D \textbf{82} (2010), 116010
doi:10.1103/PhysRevD.82.116010
[arXiv:1008.1783 [hep-ph]].

\bibitem{Fox:2011pm}
P.~J.~Fox, R.~Harnik, J.~Kopp and Y.~Tsai,
Phys. Rev. D \textbf{85} (2012), 056011
doi:10.1103/PhysRevD.85.056011
[arXiv:1109.4398 [hep-ph]].

\bibitem{Rajaraman:2011wf}
A.~Rajaraman, W.~Shepherd, T.~M.~P.~Tait and A.~M.~Wijangco,
Phys. Rev. D \textbf{84} (2011), 095013
doi:10.1103/PhysRevD.84.095013
[arXiv:1108.1196 [hep-ph]].

\bibitem{Ding:2013nvx}
R.~Ding, Y.~Liao, J.~Y.~Liu and K.~Wang,
JCAP \textbf{05} (2013), 028
doi:10.1088/1475-7516/2013/05/028
[arXiv:1302.4034 [hep-ph]].

\bibitem{Cao:2009uw}
Q.~H.~Cao, C.~R.~Chen, C.~S.~Li and H.~Zhang,
JHEP \textbf{08} (2011), 018
doi:10.1007/JHEP08(2011)018
[arXiv:0912.4511 [hep-ph]].

\bibitem{Nelson:2013pqa}
A.~Nelson, L.~M.~Carpenter, R.~Cotta, A.~Johnstone and D.~Whiteson,
Phys. Rev. D \textbf{89} (2014) no.5, 056011
doi:10.1103/PhysRevD.89.056011
[arXiv:1307.5064 [hep-ph]].

\bibitem{Crivellin:2015wva}
A.~Crivellin, U.~Haisch and A.~Hibbs,
Phys. Rev. D \textbf{91} (2015), 074028
doi:10.1103/PhysRevD.91.074028
[arXiv:1501.00907 [hep-ph]].

\bibitem{Bell:2012rg}
N.~F.~Bell, J.~B.~Dent, A.~J.~Galea, T.~D.~Jacques, L.~M.~Krauss and T.~J.~Weiler,
Phys. Rev. D \textbf{86} (2012), 096011
doi:10.1103/PhysRevD.86.096011
[arXiv:1209.0231 [hep-ph]].

\bibitem{Carpenter:2012rg}
L.~M.~Carpenter, A.~Nelson, C.~Shimmin, T.~M.~P.~Tait and D.~Whiteson,
Phys. Rev. D \textbf{87} (2013) no.7, 074005
doi:10.1103/PhysRevD.87.074005
[arXiv:1212.3352 [hep-ex]].

\bibitem{Alves:2015dya}
A.~Alves and K.~Sinha,
Phys. Rev. D \textbf{92} (2015) no.11, 115013
doi:10.1103/PhysRevD.92.115013
[arXiv:1507.08294 [hep-ph]].

\bibitem{Mao:2014rga}
M.~Song, G.~Li, W.~G.~Ma, R.~Y.~Zhang and J.~Y.~Guo,
JHEP \textbf{09} (2014), 069
doi:10.1007/JHEP09(2014)069
[arXiv:1403.2142 [hep-ph]].

\bibitem{Lopez:2014qja}
N.~Lopez, L.~M.~Carpenter, R.~Cotta, M.~Frate, N.~Zhou and D.~Whiteson,
Phys. Rev. D \textbf{89} (2014) no.11, 115013
doi:10.1103/PhysRevD.89.115013
[arXiv:1403.6734 [hep-ph]].

\bibitem{Carpenter:2013xra}
L.~Carpenter, A.~DiFranzo, M.~Mulhearn, C.~Shimmin, S.~Tulin and D.~Whiteson,
Phys. Rev. D \textbf{89} (2014) no.7, 075017
doi:10.1103/PhysRevD.89.075017
[arXiv:1312.2592 [hep-ph]].

\bibitem{Berlin:2014cfa}
A.~Berlin, T.~Lin and L.~T.~Wang,
JHEP \textbf{06} (2014), 078
doi:10.1007/JHEP06(2014)078
[arXiv:1402.7074 [hep-ph]].

\bibitem{Bartels:2012ex}
C.~Bartels, M.~Berggren and J.~List,
Eur. Phys. J. C \textbf{72} (2012), 2213
doi:10.1140/epjc/s10052-012-2213-9
[arXiv:1206.6639 [hep-ex]].

\bibitem{Chae:2012bq}
Y.~J.~Chae and M.~Perelstein,
JHEP \textbf{05} (2013), 138
doi:10.1007/JHEP05(2013)138
[arXiv:1211.4008 [hep-ph]].

\bibitem{Dreiner:2012xm}
H.~Dreiner, M.~Huck, M.~Kr\"amer, D.~Schmeier and J.~Tattersall,
Phys. Rev. D \textbf{87} (2013) no.7, 075015
doi:10.1103/PhysRevD.87.075015
[arXiv:1211.2254 [hep-ph]].

\bibitem{Yu:2013aca}
Z.~H.~Yu, Q.~S.~Yan and P.~F.~Yin,
Phys. Rev. D \textbf{88} (2013) no.7, 075015
doi:10.1103/PhysRevD.88.075015
[arXiv:1307.5740 [hep-ph]].

\bibitem{Richard:2014vfa}
F.~Richard, G.~Arcadi and Y.~Mambrini,
Eur. Phys. J. C \textbf{75} (2015), 171
doi:10.1140/epjc/s10052-015-3379-8
[arXiv:1411.0088 [hep-ex]].

\bibitem{Rossi-Torres:2015eua}
F.~Rossi-Torres and C.~A.~Moura,
Phys. Rev. D \textbf{92} (2015) no.11, 115022
doi:10.1103/PhysRevD.92.115022
[arXiv:1503.06475 [hep-ph]].

\bibitem{Habermehl:2017dxh}
M.~Habermehl, K.~Fujii, J.~List, S.~Matsumoto and T.~Tanabe,
PoS \textbf{ICHEP2016} (2016), 155
doi:10.22323/1.282.0155
[arXiv:1702.05377 [hep-ex]].

\bibitem{Gao:2017tgx}
M.~Jin and Y.~Gao,
Eur. Phys. J. C \textbf{78} (2018) no.8, 622
doi:10.1140/epjc/s10052-018-6093-5
[arXiv:1712.02140 [hep-ph]].

\bibitem{Liu:2019ogn}
Z.~Liu, Y.~H.~Xu and Y.~Zhang,
JHEP \textbf{06} (2019), 009
doi:10.1007/JHEP06(2019)009
[arXiv:1903.12114 [hep-ph]].

\bibitem{Ghosh:2019rtj}
D.~K.~Ghosh, T.~Katayose, S.~Matsumoto, I.~Saha, S.~Shirai and T.~Tanabe,
Phys. Rev. D \textbf{101} (2020) no.1, 015007
doi:10.1103/PhysRevD.101.015007
[arXiv:1906.06864 [hep-ph]].

\bibitem{Habermehl:2020njb}
M.~Habermehl, M.~Berggren and J.~List,
Phys. Rev. D \textbf{101} (2020) no.7, 075053
doi:10.1103/PhysRevD.101.075053
[arXiv:2001.03011 [hep-ex]].

\bibitem{Neng:2014mga}
N.~Wan, M.~Song, G.~Li, W.~G.~Ma, R.~Y.~Zhang and J.~Y.~Guo,
Eur. Phys. J. C \textbf{74} (2014) no.12, 3219
doi:10.1140/epjc/s10052-014-3219-2
[arXiv:1403.7921 [hep-ph]].

\bibitem{Yu:2014ula}
Z.~H.~Yu, X.~J.~Bi, Q.~S.~Yan and P.~F.~Yin,
Phys. Rev. D \textbf{90} (2014) no.5, 055010
doi:10.1103/PhysRevD.90.055010
[arXiv:1404.6990 [hep-ph]].

\bibitem{Liu:2017zdh}
J.~Liu, L.~T.~Wang, X.~P.~Wang and W.~Xue,
Phys. Rev. D \textbf{97} (2018) no.9, 095044
doi:10.1103/PhysRevD.97.095044
[arXiv:1712.07237 [hep-ph]].

\bibitem{Kadota:2018lrt}
K.~Kadota and A.~Spray,
JHEP \textbf{02} (2019), 017
doi:10.1007/JHEP02(2019)017
[arXiv:1811.00560 [hep-ph]].

\bibitem{dEnterria:2017dac}
D.~d'Enterria,
PoS \textbf{ICHEP2016} (2017), 434
doi:10.22323/1.282.0434
[arXiv:1701.02663 [hep-ex]].

\bibitem{Cotta:2012nj}
R.~C.~Cotta, J.~L.~Hewett, M.~P.~Le and T.~G.~Rizzo,
Phys. Rev. D \textbf{88} (2013), 116009
doi:10.1103/PhysRevD.88.116009
[arXiv:1210.0525 [hep-ph]].

\bibitem{Rajaraman:2012fu}
A.~Rajaraman, T.~M.~P.~Tait and A.~M.~Wijangco,
Phys. Dark Univ. \textbf{2} (2013), 17-21
doi:10.1016/j.dark.2012.12.001
[arXiv:1211.7061 [hep-ph]].

\bibitem{Chen:2013gya}
J.~Y.~Chen, E.~W.~Kolb and L.~T.~Wang,
Phys. Dark Univ. \textbf{2} (2013), 200-218
doi:10.1016/j.dark.2013.11.002
[arXiv:1305.0021 [hep-ph]].

\bibitem{Crivellin:2014gpa}
A.~Crivellin and U.~Haisch,
Phys. Rev. D \textbf{90} (2014), 115011
doi:10.1103/PhysRevD.90.115011
[arXiv:1408.5046 [hep-ph]].

\bibitem{Shifman:1978zn}
M.~A.~Shifman, A.~I.~Vainshtein and V.~I.~Zakharov,
Phys. Lett. B \textbf{78} (1978), 443-446
doi:10.1016/0370-2693(78)90481-1

\bibitem{Frandsen:2012db}
M.~T.~Frandsen, U.~Haisch, F.~Kahlhoefer, P.~Mertsch and K.~Schmidt-Hoberg,
JCAP \textbf{10} (2012), 033
doi:10.1088/1475-7516/2012/10/033
[arXiv:1207.3971 [hep-ph]].

\bibitem{Tang:2015uha}
Y.~L.~Tang, C.~Zhang and S.~h.~Zhu,
Phys. Rev. D \textbf{94} (2016) no.1, 011702
doi:10.1103/PhysRevD.94.011702
[arXiv:1508.01095 [hep-ph]].

\bibitem{Han:2018rkz}
C.~Han, R.~Li, R.~Q.~Pan and K.~Wang,
Phys. Rev. D \textbf{98} (2018) no.11, 115003
doi:10.1103/PhysRevD.98.115003
[arXiv:1802.03679 [hep-ph]].

\bibitem{Degrande:2011ua}
C.~Degrande, C.~Duhr, B.~Fuks, D.~Grellscheid, O.~Mattelaer and T.~Reiter,
Comput. Phys. Commun. \textbf{183} (2012), 1201-1214
doi:10.1016/j.cpc.2012.01.022
[arXiv:1108.2040 [hep-ph]].

\bibitem{Alloul:2013bka}
A.~Alloul, N.~D.~Christensen, C.~Degrande, C.~Duhr and B.~Fuks,
Comput. Phys. Commun. \textbf{185} (2014), 2250-2300
doi:10.1016/j.cpc.2014.04.012
[arXiv:1310.1921 [hep-ph]].

\bibitem{Alwall:2014hca}
J.~Alwall, R.~Frederix, S.~Frixione, V.~Hirschi, F.~Maltoni, O.~Mattelaer, H.~S.~Shao, T.~Stelzer, P.~Torrielli and M.~Zaro,
JHEP \textbf{07} (2014), 079
doi:10.1007/JHEP07(2014)079
[arXiv:1405.0301 [hep-ph]].

\bibitem{Sjostrand:2006za}
T.~Sjostrand, S.~Mrenna and P.~Z.~Skands,
JHEP \textbf{05} (2006), 026
doi:10.1088/1126-6708/2006/05/026
[arXiv:hep-ph/0603175 [hep-ph]].

\bibitem{Ovyn:2009tx}
S.~Ovyn, X.~Rouby and V.~Lemaitre,
[arXiv:0903.2225 [hep-ph]].

\bibitem{AbelleiraFernandez:2012cc}
J.~L.~Abelleira Fernandez \textit{et al.} [LHeC Study Group],
J. Phys. G \textbf{39} (2012), 075001
doi:10.1088/0954-3899/39/7/075001
[arXiv:1206.2913 [physics.acc-ph]].

\bibitem{CEPCStudyGroup:2018rmc}
 [CEPC Study Group],
[arXiv:1809.00285 [physics.acc-ph]].

\bibitem{CEPCStudyGroup:2018ghi}
J.~B.~Guimar\~aes da Costa \textit{et al.} [CEPC Study Group],
[arXiv:1811.10545 [hep-ex]].

\bibitem{Aad:2020arf}
G.~Aad \textit{et al.} [ATLAS],
JHEP \textbf{02} (2021), 226
doi:10.1007/JHEP02(2021)226
[arXiv:2011.05259 [hep-ex]].

\bibitem{Aad:2021egl}
G.~Aad \textit{et al.} [ATLAS],
Phys. Rev. D \textbf{103} (2021) no.11, 112006
doi:10.1103/PhysRevD.103.112006
[arXiv:2102.10874 [hep-ex]].

\bibitem{Weiner:2012cb}
N.~Weiner and I.~Yavin,
Phys. Rev. D \textbf{86} (2012), 075021
doi:10.1103/PhysRevD.86.075021
[arXiv:1206.2910 [hep-ph]].

\bibitem{Belanger:2013oya}
G.~Belanger, F.~Boudjema, A.~Pukhov and A.~Semenov,
Comput. Phys. Commun. \textbf{185} (2014), 960-985
doi:10.1016/j.cpc.2013.10.016
[arXiv:1305.0237 [hep-ph]].

\bibitem{Zyla:2020zbs}
P.~A.~Zyla \textit{et al.} [Particle Data Group],
PTEP \textbf{2020} (2020) no.8, 083C01
doi:10.1093/ptep/ptaa104

\bibitem{XENON:2017vdw}
E.~Aprile \textit{et al.} [XENON],
Phys. Rev. Lett. \textbf{119} (2017) no.18, 181301
doi:10.1103/PhysRevLett.119.181301
[arXiv:1705.06655 [astro-ph.CO]].

\bibitem{XENON:2018voc}
E.~Aprile \textit{et al.} [XENON],
Phys. Rev. Lett. \textbf{121} (2018) no.11, 111302
doi:10.1103/PhysRevLett.121.111302
[arXiv:1805.12562 [astro-ph.CO]].

\bibitem{XENON:2023sxq}
E.~Aprile \textit{et al.} [XENON],
Phys. Rev. Lett. \textbf{131} (2023) no.4, 041003
doi:10.1103/PhysRevLett.131.041003
[arXiv:2303.14729 [hep-ex]].

\bibitem{PandaX-4T:2021bab}
Y.~Meng \textit{et al.} [PandaX-4T],
Phys. Rev. Lett. \textbf{127} (2021) no.26, 261802
doi:10.1103/PhysRevLett.127.261802
[arXiv:2107.13438 [hep-ex]].

\bibitem{Planck:2018vyg}
N.~Aghanim \textit{et al.} [Planck],
Astron. Astrophys. \textbf{641} (2020), A6
[erratum: Astron. Astrophys. \textbf{652} (2021), C4]
doi:10.1051/0004-6361/201833910
[arXiv:1807.06209 [astro-ph.CO]].

\bibitem{Ruppin:2014bra}
F.~Ruppin, J.~Billard, E.~Figueroa-Feliciano and L.~Strigari,
Phys. Rev. D \textbf{90} (2014) no.8, 083510
doi:10.1103/PhysRevD.90.083510
[arXiv:1408.3581 [hep-ph]].

\bibitem{Fermi-LAT:2015att}
M.~Ackermann \textit{et al.} [Fermi-LAT],
Phys. Rev. Lett. \textbf{115} (2015) no.23, 231301
doi:10.1103/PhysRevLett.115.231301
[arXiv:1503.02641 [astro-ph.HE]].

\end{thebibliography}
\end{document}